\documentclass[twocolumn]{aastex631}
\usepackage{amssymb,amsmath}

\newcommand{\muller}{Mueller }
\newcommand{\mat}[1]{\ensuremath{\mathbb{#1}}}

\newcommand{\abs}[1]{\ensuremath{\left\lvert #1 \right\rvert}}

\newcommand{\functional}[1]{\ensuremath{\mathcal{#1}}} 
\newcommand{\minfunc}{\ensuremath{\functional{L}}} 
\newcommand{\mindia}{\ensuremath{\minfunc}_P} 
\newcommand{\minret}{\ensuremath{\minfunc}_R} 

\accepted{April 5, 2021}

\submitjournal{\apj}

\shorttitle{Model-based Ad Hoc X-Talk}
\shortauthors{Jaeggli et al.}

\graphicspath{{./}{figures/}}

\begin{document}

\title{A Model-based Technique for Ad Hoc Correction of Instrumental \\ Polarization in Solar Spectropolarimetry}

\correspondingauthor{Sarah A. Jaeggli}
\email{sjaeggli@nso.edu}

\author[0000-0001-5459-2628]{Sarah A. Jaeggli}
\affiliation{National Solar Observatory \\
22 Ohia Ku Street \\
Pukalani, HI 96768, USA}

\author[0000-0002-7451-9804]{Thomas A. Schad}
\affiliation{National Solar Observatory \\
22 Ohia Ku Street \\
Pukalani, HI 96768, USA}

\author[0000-0002-8259-8303]{Lucas A. Tarr}
\affiliation{National Solar Observatory \\
22 Ohia Ku Street \\
Pukalani, HI 96768, USA}

\author[0000-0002-3215-7155]{David M. Harrington}
\affiliation{National Solar Observatory \\
22 Ohia Ku Street \\
Pukalani, HI 96768, USA}

\begin{abstract}

We present a new approach for correcting instrumental polarization by modeling the non-depolarizing effects of a complex series of optical elements to determine physically realizable \muller matrices.  Provided that the \muller matrix of the optical system can be decomposed into a general elliptical diattenuator and general elliptical retarder, it is possible to model the cross-talk between both the polarized and unpolarized states of the Stokes vector and then use the acquired science observations to determine the best-fit free parameters.  Here, we implement a minimization for solar spectropolarimetric measurements containing photospheric spectral lines sensitive to the Zeeman effect using physical constraints provided by polarized line and continuum formation.  This model-based approach is able to provide an accurate correction even in the presence of large amounts of polarization cross-talk and conserves the physically meaningful magnitude of the Stokes vector, a significant improvement over previous ad hoc techniques.

\end{abstract}

\keywords{Spectropolarimetry (1973) --- Calibration (2179) --- Solar physics (1476) --- Solar magnetic fields (1503)}


\section{Introduction}\label{sec:intro}
In the field of solar physics, spectropolarimetry at optical wavelengths is a critical tool used to infer the magnetic field vector from the photosphere to chromosphere and corona.  
The Zeeman effect, Paschen-Bach effect, Hanle effect, atomic level polarization, and scattering polarization produce polarized signatures in spectral lines as well as continuum emission in the case of scattering polarization \citep{landi04}.
Accurate measurements of linear and circular polarization and the total intensity of light are critical for recovery of the strength and direction of the magnetic field along with other diagnostics of the solar plasma using spectropolarimetric inversions, which are now mainstays of solar physics \citep{deltoro16}.
There are many instruments designed specifically for the measurement of these polarized diagnostics.  \citet{iglesias19} give an extensive list of recently operating and upcoming instruments.

Instrument and telescope optical systems cause exchange between polarized and unpolarized light, and between linear and circular polarization states.
Reflection from surfaces, transmission through media, and interaction with coatings can modify the polarization state of light in different ways dependent on wavelength and angle of incidence.
Astronomical instruments often require changes in optical geometry during observations to maintain tracking and orientation of the field of view, causing changes in the polarization response of the optical system.  
Telescope mirrors may be recoated to improve transmission every few years, and the polarization performance of optical elements may change after recoating \citep[see ][]{jatis7}, although detailed measurements show that even unprotected aluminum coatings are stable after initial buildup of the oxide layer immediately after deposition \citep{vanharten09}.

There are various approaches to polarimetric calibration that can combine theoretical modeling, measurements using calibration optics, direct measurements of sources with well known or assumed polarization properties, and semi-empirical modeling \citep[see, e.g.][]{jatis1, elmore92, collados03, harrington11, beck05}.
However, it is not always possible to have a well known or well calibrated telescope and instrument system.  
Polarization cross-talk may remain even after calibration measurements and polarization models have been applied due to uncertainty in the parameters of the calibration optics or uncorrected sections of the optical path, for example.

No single method for polarization calibration is without challenges, and it is desirable to have an additional method to diagnose and remove residual instrumental polarization or cross-validate results.  
There are several previously published approaches for the removal of residual instrumental polarization from spectropolarimetry that make use of the physical properties in the solar observations themselves \citep[\textit{i.e.} ad hoc corrections,][]{november91, sanchez92, kuhn94, schlichenmaier02, collados03, derks18}.  
The primary goal of these techniques is to determine and correct cross-talk between linear and circular polarization states. 
Some of these techniques also remove the cross-talk between polarized and unpolarized states, this technique is stated clearly in \citet{sanchez92}.

These ad hoc correction techniques take advantage of Zeeman-split photospheric absorption lines which are considered to form in a state near local thermodynamic equilibrium and, in the ideal case, produce very well understood line profiles \citep{unno56, landi98}.  
Although there are a few exceptions \citep[i.e.][]{derks18}, the commonly used spectral lines display the anomalous Zeeman effect, which produces a three component line profile.
For a magnetic field that is transverse with respect to the observer's line of sight, the line will show a linearly polarized profile, symmetric about the line center, with a central component and wings polarized at $90^\circ$ with respect to each other.  
For a magnetic field oriented along the observer's line of sight, the line will show a profile with circular polarization, anti-symmetric about the line center, with one wing showing left circular polarization and the other showing right circular polarization.  
Intermediate orientations of the magnetic field will show a combination of linearly and circularly polarized profiles with these same symmetry properties.

As a first step, the techniques of \citet{sanchez92}, \citet{kuhn94}, and \citet{schlichenmaier02} assume that the continuum polarization is zero and remove the cross-talk from the unpolarized intensity state into the polarized states based on their continuum values.  \citet{collados03} makes no explicit assumption about the continuum polarization, but a similar correction would be necessary unless the intensity cross-talk was negligible.
The methods then exploit the symmetry and anti-symmetry properties of Zeeman-split lines to determine the amount of polarization exchanged by the optical system assuming that on average the Sun produces ideal polarized profiles.  
\citet{sanchez92} assume that the circularly polarized profile is symmetric about the line core, while \citet{collados03} assumes that the net circular polarization is equal to zero.  
\citet{kuhn94} incorporate feature-specific information by assuming the unshifted central component of the line should not show any circular polarization signal in cases where the separation of the three components of the Zeeman-split line is very large, \textit{i.e.} in a sunspot umbra.  
All four methods assume that linear and circular polarization signals are statistically uncorrelated over the line profile.  
Above all, these techniques assume that the amount of polarization cross-talk is small.

Each of the above methods solves a linear system of equations describing the transformation from linear to circular polarization and from circular to linear polarization using some optimization technique. 
Crucially, all the above methods find the coefficients of the circular-to-linear transform as if they were independent from the linear-to-circular transform.
On the contrary, physical models for the polarization cross-talk induced by  different  kinds  of  optical  elements  show  that  the cross-talk terms between polarization states and to unpolarized light are not independent, but instead typically linked by some simple functional form.

In this paper we present a model-based approach for characterizing and correcting instrumental polarization that builds on previous ad hoc techniques.
Crucially, our model is based on the fact that any linear combination of non-depolarizing optical elements can be represented as an equivalent system composed of just an elliptical diattenuator and an elliptical retarder.
The reduced parameter set fully embodies the physics of polarized transfer within the optical system and therefore provides a more robust correction compared to the previous methods. 
In particular, our method performs well even in the presence of the high levels of polarization cross-talk that are often found in real telescope systems.
The polarization modeling, measurement, and calibration efforts for the Daniel K. Inouye Solar Telescope have demonstrated that the polarization properties of optics are well described using only the few parameters that describe the diattenuation and retardance of the reflective or transmissive optic \citep{jatis1, jatis5}.
Those efforts provide strong real-world experimental validation for our method, in addition to its sound theoretical underpinnings.

The rest of the paper is outlined as follows.
In Section \ref{sec:theory} we provide motivation for the physical approach we have selected.
In Section \ref{sec:implementation} we present our algorithm for determining the parameters.
In Section \ref{sec:application} we demonstrate our technique by using a pristine dataset with a known level of polarization cross-talk contamination.  To put our ad hoc technique in context with those of previous authors, we also apply a combination of techniques from \citet{sanchez92} and \citet{kuhn94}.
In Section \ref{sec:results} we discuss the results of our technique and explore some weaknesses of our technique and the previous ad hoc techniques.
In Section \ref{sec:end} we provide conclusions and discuss further applications for the model-based approach.

\section{Approach}\label{sec:theory}
In the Stokes-\muller formalism, any polarized state of light can be represented by the 4-element Stokes vector.
\begin{equation}
    \vec{S} = 
    \left[ {\begin{array}{c}
    I \\
    Q \\
    U \\
    V \\
    \end{array} } \right]
\end{equation}
$I$ is the total of all polarized and unpolarized light.  $Q$ and $U$ represent the linear polarization.  Positive $Q$ is oriented at $0^\circ$ or $180^\circ$ from a reference axis normal to the propagation direction while negative $Q$ is oriented at $90^\circ$ or $270^\circ$.  Positive $U$ is oriented at $45^\circ$ or $225^\circ$ while negative $U$ is oriented at $135^\circ$ or $315^\circ$. $V$ represents the circular polarization, where right hand (counter clockwise) circular polarization is positive and left hand (clockwise) circular polarization is negative.

A linear transformation of a Stokes vector is represented by a $4\times4$ \muller matrix which carries some initial vector $\vec{S}$ to a transformed vector $\vec{S}^\prime$: 
\begin{equation}
    \vec{S}^\prime = \mat{M}\vec{S}\label{eq:muellertransform}
\end{equation}
where
\begin{equation}
    \mat{M} =
    \left[ {\begin{array}{cccc}
    m_{00} & m_{10} & m_{20} & m_{30} \\
    m_{01} & m_{11} & m_{21} & m_{31} \\
    m_{02} & m_{12} & m_{22} & m_{32} \\
    m_{03} & m_{13} & m_{23} & m_{33} \\
    \end{array} } \right].
\end{equation}
It is possible to represent many kinds of optical processes using a \muller matrix, including rotation, retardance, diattenuation, and depolarization.  Extensive examples are given in \citet{chipmanbook}.  
There are a variety of tests determine if any given matrix represents a physically admissible collection of optical elements \citep[e.g.][]{kostinski93, givens93}.

The total \muller matrix for a series of optical elements can be constructed by the matrix product of the individual \muller matrices for each optical element ($\mat{M}_i$) and rotation \muller matrices ($\mat{R}_i$) that are used to  change the Stokes coordinate frame between elements, for example:
\begin{equation}
    \mat{M}_{total} = \mat{R}_3 \mat{M}_3 \mat{R}_{23} \mat{M}_2 \mat{R}_{12} \mat{M}_1 \mat{R}_1
\end{equation}
The new \muller matrix is built from right to left as each new matrix operates on the Stokes vector resulting from the operations done before. In general, these operations do not commute but are associative.

The polarization properties of optics found in telescopes and other imaging systems are typically well described by three physical processes:  reflectance or transmittance, diattenuation, and retardance.  
Non-uniform scattering and/or coatings, as well as powered optics, can introduce depolarization, but the associated 9 degrees of freedom are typically small and often ignored \citep{chipmanbook,jatis7}.  
Individual optics are subsequently modeled very well with only diattenuation and retardance, although they may have complex behavior as a function of wavelength, spatial dependence over the optic, or dependence on the angle of incidence \citep[e.g][]{jatis5}.
These are ``non-depolarizing'' \muller matrices and they have special properties.
Any combination of non-depolarizing \muller matrices are also non-depolarizing \citep{gil86}, and therefore, a series of optical elements composed of many optics of this type can still be modeled with only diattenuation and retardance.

Just as it is possible to produce a single \muller matrix through matrix multiplication, it is possible to decompose an arbitrary \muller matrix into separate \muller matrices.  Non-depolarizing \muller matrices can be decomposed into two parts: a general elliptical retarder and a general elliptical diattenuator \citep{gil16}, where the decomposition can be done in either order:
\begin{equation}\label{eqn:decomp}
    \mat{M}_{sys} = \mat{M}_{P} \mat{M}_{R} = \mat{M}_{R} \mat{M}_{D}
\end{equation}
$\mat{M}_P$ and $\mat{M}_D$ denote the \muller matrix for a left-equivalent and right-equivalent diattenuator, respectively, and $\mat{M}_R$ is the \muller matrix for a general elliptical retarder.  
$\mat{M}_P$ and $\mat{M}_D$ are related as $\mat{M}_P$  = $\mat{M}_R \mat{M}_D \mat{M}_R^{T}$.  
Below, we only use the left-equivalent diattenuator form of this decomposition.  
The matrix for a general diattenuator from \citet{chipmanbook}, normalized and expressed as a parallel decomposition, is:
\begin{align}\label{eqn:diattenuator}
    \mat{M}_P = 
    & \left[ {\begin{array}{cccc}
    1 & d_{H} & d_{45} & d_{R} \\
    d_{H} & A & 0 & 0 \\
    d_{45} & 0 & A & 0 \\
    d_{R} & 0 & 0 & A \\
    \end{array} } \right] + \nonumber \\
    & \frac{1-A}{D^2} \left[ {\begin{array}{cccc}
    0 & 0 & 0 & 0 \\
    0 & d_{H}^2 & d_{45} d_{H} & d_{R} d_{H} \\
    0 & d_{H} d_{45} & d_{45}^2 & d_{R} d_{45} \\
    0 & d_{H} d_{R} & d_{45} d_{R} & d_{R}^2 \\
    \end{array} } \right]
\end{align}
where we can also write the normalized diattenuation vector $[d_{H},d_{45},d_{R}]^{T}$ in equivalent spherical coordinates:
\begin{align}
    d_{H} & =  D \cos{\alpha} \sin{\beta} \\
    d_{45} & =  D \sin{\alpha} \sin{\beta} \\
    d_{R} & =  D \cos{\beta} \\
    A & =  \sqrt{1 - D^2}
\end{align}
where $D$ is the magnitude of the diattenuation vector ($0 \le D \le 1$) and $\alpha$ and $\beta$ are the equivalent angles for the vector.

An elliptical retarder acts a generalized rotation of a Stokes vector on the Poincar\'e sphere.  Its \muller matrix is most often written to include an axis-angle rotation matrix where the eigenpolarization direction (or fast axis) defines the axis of rotation \cite[see Equation 29 of][]{chipmanbook}.  For our purposes, it is advantageous to convert this matrix into a combination of more simple components.  \citet{manhas06}, for example, shows that an elliptical retarder can be equivalently written as the product of a linear retarder with a fast axis angle oriented at some angle and a rotation matrix.  Here we employ the extrinsic z-x-z convention for Euler angles, which corresponds to the product of \muller matrices for a circular retarder, a linear retarder with fast axis at $0^\circ$, and another circular retarder:
\begin{equation}\label{eqn:retarder}
    \mat{M}_{R} = \mat{M}_{R3} \mat{M}_{R2} \mat{M}_{R1}
\end{equation}
where
\begin{align} 
    \mat{M}_{R1} = & \begin{bmatrix}
    1 & 0 & 0 & 0 \\
    0 & \cos{\phi} & \sin{\phi} & 0 \\
    0 & -\sin{\phi} & \cos{\phi} & 0 \\
    0 & 0 & 0 & 1
    \end{bmatrix} \label{eq:mr1}\\
    \mat{M}_{R2} = & \begin{bmatrix}
    1 & 0 & 0 & 0 \\
    0 & 1 & 0 & 0 \\
    0 & 0 & \cos{\delta} & \sin{\delta} \\
    0 & 0 & -\sin{\delta} & \cos{\delta} 
    \end{bmatrix}\label{eq:mr2} \\
    \mat{M}_{R3} = & \begin{bmatrix}
    1 & 0 & 0 & 0 \\
    0 & \cos{\theta} & \sin{\theta} & 0 \\
    0 & -\sin{\theta} & \cos{\theta} & 0 \\
    0 & 0 & 0 & 1 
    \end{bmatrix}.\label{eq:mr3}
\end{align} 
Now that we've broken down the optical model into relatively simple parts, we can see how they can be applied to the problem of solar spectropolarimetry in the next section.

\section{Implementation}\label{sec:implementation}
Let's assume we have a spectropolarimetric measurement from an instrument with two spatial dimensions and one spectral dimension ($x$, $y$, and $\lambda$ respectively).  The measurement contains an area of the Sun with strong magnetic fields with different orientations.  The original Stokes spectra from this region, $\vec{S}_{Orig} = \left[ I_{Orig}, Q_{Orig}, U_{Orig}, V_{Orig} \right]^T$, are produced by photospheric spectral lines sensitive to the Zeeman effect, and consist of symmetric Stokes Q and U profiles and anti-symmetric Stokes V profiles as well as Stokes I.  The demodulation of the observed intensities has been done to retrieve the measured Stokes vector, $\vec{S}_{Meas} = \left[ I_{Meas}, Q_{Meas}, U_{Meas}, V_{Meas} \right]^T$, but there is still some uncorrected optical path that mixes up the polarization and intensity signals of $\vec{S}_{Orig}$; that is, there is an unknown \muller matrix that transforms $\vec{S}_{Orig}$ into $\vec{S}_{Meas}$ as in Equation \ref{eq:muellertransform}.  We assume that the instrument geometry is effectively static during the observation so that the rotation matrices $\mat{R}_i$ are fixed.  We also assume that the instrument polarization properties are constant over the observed wavelengths and field-of-view of the measurement.
This last assumption is made primarily to keep the current exposition as clear as possible.  Field of view and wavelength dependent polarization may be significant and our technique could be extended to account for their effects.  We discuss this more in Section \ref{sec:end}.

Our goal is to determine the diattenuator ($\mat{M}_P$) and retarder ($\mat{M}_R$) \muller matrices that minimize certain criteria based on physical assumptions about the polarized signals from the Sun.  The diattenuator and retarder combine to produce the recovered, general \muller matrix $\mat{M}_{Rec}$.
We discuss different minimizing functionals $\mathcal{L}$ below.  The recovered Stokes vector is produced by applying the inverse of the recovered \muller matrix to the measured Stokes vector during iterative minimization steps:
\begin{equation}
    \vec{S}_{Rec} = \mat{M}_{Rec}^{-1} \vec{S}_{Meas}.
\end{equation}
After the minimization, ideally $\vec{S}_{Rec} \simeq \vec{S}_{Orig}$.

The determination of the diattenuation matrix may be complicated by the presence of polarization in the continuum of the Sun, which is linearly polarized along the direction perpendicular to the limb due to scattering in the solar atmosphere.  The magnitude of linear polarization increases from disk center to the limb and is dependent on wavelength \citep{fluri99}.  
For small fields of view (on the scale of sunspots) it is not possible to distinguish linear polarization of the continuum from diattenuation induced by the optical system.
To accurately recover diattenuation parameters, we need to restrict ourselves to regions near disk center where we can assume that the continuum is unpolarized on average.  
Extension of this technique to regions with a linearly polarized continuum should be possible provided that the continuum polarization can be accurately modeled.  This is discussed further in Section \ref{sec:end}.

Because we have assumed that the continuum is unpolarized, and Stokes I is only exchanged with Q, U, or V by diattenuation, the determination of diattenuation and retardance terms can be separated into two different steps.  
The main reason to separate the models is a practical one:  
spectropolarimetric data often require destreaking in the spatial domain of Stokes Q, U, and V.  This involves subtracting any residual Stokes I cross-talk from the spectrum at each $(x,y)$ position, e.g. for Stokes Q this might look like $Q_{xy\lambda}-I_{xy\lambda}\left\langle Q_{xy\lambda}/I_{xy\lambda}\right\rangle_{\lambda_\text{cont.}}$ where the term in brackets is an average over continuum wavelengths.
Such streaks might arise from slight vertical mismatch in the beam registration for a dual-beam spectopolarimeter or other instrument instabilities.  
The destreaking step needs to be done after removal of cross-talk between Stokes I and the polarized states; and, to ensure streaks are not interpreted as polarized cross-talk, they need to be removed before deriving the retardance terms.  This was the motivation for selecting the left-equivalent diattenuator decomposition in Equation~\ref{eqn:decomp}.

Armed with what we know about Stokes I, Q, U, and V, we can now construct a minimization criteria that reduces the diattenuation cross-talk.  We want to choose a criteria that minimizes the correlation of Stokes Q, U, and V to I.  To achieve this, we take the inner product of Stokes I on Q, U, and V, integrated over wavelength, and summed over all spatial positions:
\begin{equation}\label{eqn:min1}
    \mindia=\sum\limits_{xy} \Bigl( 
    {\Bigl\vert \sum\limits_{\lambda} IQ \Bigr\vert} + 
    {\Bigl\vert \sum\limits_{\lambda} IU \Bigr\vert} + 
    {\Bigl\vert \sum\limits_{\lambda} IV \Bigr\vert} 
    \Bigr)
\end{equation}
where the $I$, $Q$, $U$, and $V$ terms are the ones produced by application of the recovered diattenuation \muller matrix $\mat{M}_{P Rec}^{-1}\vec{S}_{Meas}$.
This has the effect of both minimizing the continuum polarization and reducing the correlation of the uniformly signed Stokes I line profiles against the Stokes Q, U, and V profiles which have both positive and negative sign.
In preparation for the next step, we apply the inverse of the final diattenuation \muller matrix from the minimization to the measured data to produce a Stokes vector with Stokes I cross-talk removed ($\vec{S}_{P Rec}$).

Now we can turn our attention to the problem of cross-talk between only the polarized components of the Stokes vector.
Let us consider how the three matrices in Equation \ref{eqn:retarder} exchange polarization of the original Zeeman-split spectral profiles from the Sun, where the profiles from Q and U are symmetric and the V profile is anti-symmetric in wavelength about line center.
The first matrix $\mat{M}_{R1}$ exchanges Q and U.  This is equivalent to a rotation of $\phi/2$ in the plane of the sky, where the factor of 2 recognizes that the Q and U axes are only 45 degrees apart in the Cartesian frame fixed to the Sun.  Q and U being similar, symmetric kinds of profiles, we cannot tell that they have been exchanged simply by looking at the profile shapes.
The second matrix $\mat{M}_{R2}$ exchanges U and V signals.  This exchange manifests, quite obviously, as anti-symmetric line profiles mixed into U and symmetric profiles mixed into V.
The final matrix $\mat{M}_{R3}$ exchanges the new U profile (containing V signal) and the Q profile resulting from the first matrix multiplication (that still looks symmetric).  This rotation is also obvious because anti-symmetric profiles are now mixed with the symmetric Q profiles.  

The first transformation by $\mat{M}_{R1}$ cannot be distinguished based on the profile shape alone and must be determined by another method.  A minimization based on profile shapes can only act on the $\mat{M}_{R3}\mat{M}_{R2}$ portion of Equation \ref{eqn:retarder}.  
We construct a single model from these two matrices and use a minimization that enforces zero net circular polarization (Stokes V from the Sun should be anti-symmetric) and minimizes the correlation of Q to V and U to V (the product of symmetric and anti-symmetric functions is anti-symmetric).  The minimizing functional is
\begin{equation}\label{eqn:min2}
    \minret=\sum\limits_{xy} \Bigl( 
    {\Bigl\vert \sum\limits_{\lambda} QV \Bigr\vert} + 
    {\Bigl\vert \sum\limits_{\lambda} UV \Bigr\vert} + 
    {\bigl( \sum\limits_{\lambda} V \bigr)^2} 
    \Bigr),
\end{equation}
where the Q, U, and V terms are the ones produced by application of the recovered retardance \muller matrix $\mat{M}_{R23 Rec}\vec{S}_{P Rec}$.
There are several degenerate combinations of the angles $\theta$ and $\delta$ appearing in Equations \ref{eq:mr1}-\ref{eq:mr3} that satisfy the minimization criteria $\mathcal{L}_R\rightarrow 0$.  
There is no way to distinguish between positive and negative sign in Q, U, or V, or to tell to what degree Q and U have been exchanged, but by knowing a single solution we can easily determine the others.  
The final \muller matrix from the minimization, $\mat{M}_{R23 Rec}$, can be applied at this point to recover the Stokes vector with the combined cross-talk between Stokes Q and U and Stokes V removed, $\vec{S}_{R23 Rec}$.

Several aspects of the last result require additional information.
The $\mat{M}_{R1}$ matrix in Equation \ref{eqn:retarder} that specifies the orientation of the linear polarization coordinate system on the Sun, and the sign degeneracies in the recovered Stokes V, still remain be determined and cannot be resolved using the minimization techniques described above.
There are several ways in which these issues might be resolved:
\begin{enumerate}
    {\item[(\textit{i})] Approximate knowledge of the instrument \muller matrix, based either on direct polarization measurements or modeling of the optical properties of each element, could help resolve the ambiguities and retrieve the linear polarization coordinate system.}
    {\item[(\textit{ii})] The sign convention and rotation could be determined self-consistently using the expected geometry of the solar magnetic field with a technique similar to those used to resolve the the $180^\circ$ ambiguity for magnetic fields \citep[e.g.][]{metcalf94}.}
    {\item[(\textit{iii})]  The corrected Stokes profiles resulting above could be compared to an external, fully calibrated telescope with known or small polarization errors.}
\end{enumerate}
If \textit{ii} has been done, then all that remains to be determined in \textit{iii} is the sign of the magnetic field in a global sense, either into or out of the Sun.  This can be done by comparison to a line-of-sight magnetogram.

Once they are determined, the results of these corrections can be used to retrieve the instrument \muller matrix at the time of the measurement.  All of the pieces, the diattenuation matrix and the retarder matrix with the sign and angle ambiguities resolved, are necessary to reconstruct the proper \muller matrix for the optical system.

\section{Application}\label{sec:application}

\subsection{Dataset}
\begin{figure*}
    \centering
    \includegraphics[width=7.5in]{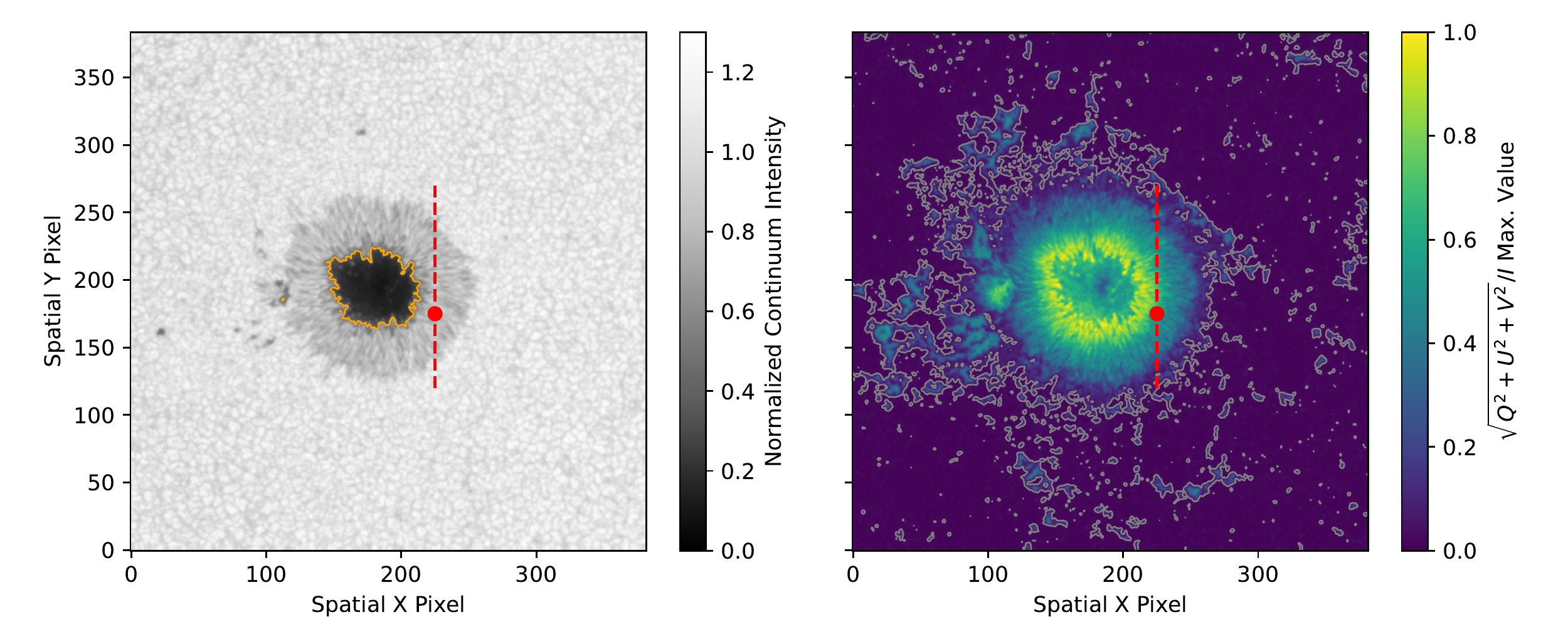}
    \caption{Raster maps showing the continuum intensity (left panel) and the maximum of the fractional polarization ($\sqrt{Q^2+U^2+V^2}/I$) in the Fe I 6302.5 \AA\ line (right panel).  The gray contours in the right hand panel separates regions with large and small polarization signatures at the 5\% level selected as input to our technique and for determination of Stokes V to Q and U cross-talk coefficients for the \citet{kuhn94} method.  The orange contour in the left hand panel shows the umbral region selected for determining Stokes Q and U to V cross-talk coefficients for the \citet{kuhn94} method. The spectra from the slice indicated by the red dashed line are shown in Figure \ref{fig:spectra}, while the individual spectral profiles for the red point are shown in Figure \ref{fig:profiles}.}
    \label{fig:maps}
\end{figure*}

To demonstrate our technique, we have selected a dataset from the Solar Optical Telescope Spectropolarimeter (SOT/SP) on board the Japanese Space Agency's {\it Hinode} spacecraft \citep{kosugi07}.  The polarization properties of SOT/SP were rigorously calibrated before launch at the $10^{-3}$ level relative to the Stokes I intensity \citep{ichimoto08}, and as a space-based instrument it provides very stable observations with high signal-to-noise.
These properties make Hinode a useful tool for resolving degeneracies by method \textit{iii}, above.  More information about the instrument can be found in \citet{lites13a}.  

SOT/SP observes a 2.4 \AA\ wide bandpass centered at 6302 \AA\ that contains two strong photospheric \ion{Fe}{1} lines at 6301.5 and 6302.5 \AA\ sensitive to the Zeeman effect with effective Land\'e g factors of 1.67 and 2.50 respectively.  This bandpass also includes several weaker lines that appear more strongly in sunspots.  These lines produce a low level of additional polarization in and around the stronger \ion{Fe}{1} lines.

The selected dataset is a raster scan of the active region NOAA 11092 starting at 2010-08-03 15:00:53 UT when the main sunspot was nearest the central meridian of the Sun.  
NOAA 11092 was a classic $\alpha$-type active region in the Hale classification \citep{hale19}, with a large monolithic sunspot in the leading polarity that was stable over many days.  There was no significant flare activity recorded on this day, or on the days before or after the observation.  
The raster took approximately 25 minutes and covered an area of 121$\times$122 arcsec$^2$ including the sunspot and surrounding active network.  The SP fast mode was used for this scan, resulting in a spatial sampling of 0.3 arcsec in the directions parallel and perpendicular to the slit.

We obtained the Level 1 SOT/SP data from the High Altitude Observatory's Community Spectropolarimetric Analysis Center (\url{https://www2.hao.ucar.edu/csac}).  Level 1 data includes the Stokes I, Q, U, and V spectra for each step in the raster.  This data has already been reduced and corrected for a variety of effects including the removal of the instrument \muller matrix using the \verb+SP_PREP+ calibration package \citep{lites13b}.   These routines apply a final polarimetric correction that removes any residual continuum polarization from the spectra at every spatial position in Stokes Q and U.  However, the destreaking of the Stokes V spectra is only performed for spatial positions where the Stokes V signal falls below a certain threshold because the wing of the profile can extend to the edges of the bandpass in regions with strong magnetic fields \citep[see Section 3.2.9 of][]{lites13b}.  In the dataset we have selected it does not appear that destreaking has been applied to Stokes V at all.

The subsequent data processing and analysis was performed using the {\it Python 3} programming language making use of the {\it NumPy}, {\it SciPy}, {\it Astropy}, and a few other publicly available modules\footnote{{\it IPython} notebooks containing this analysis are available on \url{https://github.com/sajaeggli/adhoc_xtalk}.}.  
To determine the intensity normalization to the quiet-Sun continuum, we constructed a histogram of Stokes I using 200 bins and took the intensity value at the peak of histogram.  Because the majority of the raster is over granulation and most of the spectral pixels are in the continuum, the value at this peak should be representative of the average quiet-Sun continuum intensity.  We normalized the Stokes components at each spectral and spatial pixel by this value.  
We then applied destreaking to Stokes V using the same technique as describe by \citet{lites13b} but using different threshold values: we apply destreaking only for spatial locations where the mean value of $\abs{V/I}$ is less than 8\% and then use the portions of the Stokes V spectrum with values of less than 1\% in $\abs{V/I}$.

\begin{figure*}
    \centering
    \includegraphics[width=6.25in]{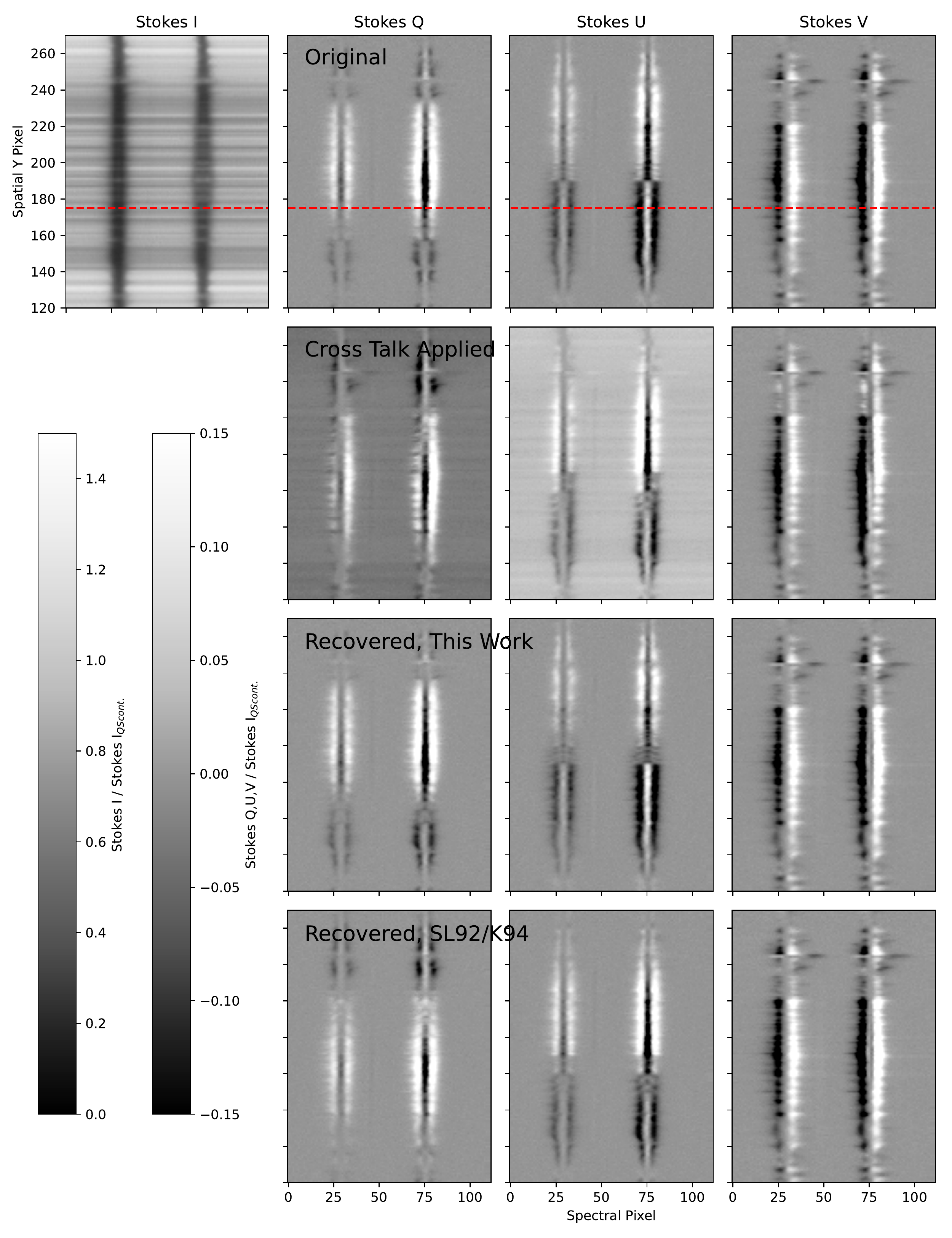}
    \caption{An example from the position indicated by the dashed red line in Figure \ref{fig:maps} showing the original Stokes spectra (top row), after application of the cross-talk \muller matrix (second row), the recovered Stokes spectra using the technique in this work (third row), and the recovered Stokes spectra using the SL92/K94 technique (bottom row).  Stokes I is the same in each case and is shown only for the top row.  The color scale for the Stokes Q, U, and V components is saturated at $\pm$0.15 but values go up to about $\pm$0.3.  The spectral profiles from the location with the red dashed line are shown in Figure \ref{fig:profiles}.}
    \label{fig:spectra}
\end{figure*}

Figure \ref{fig:maps} shows a map of the continuum intensity (left panel) and the maximum value of the fractional polarization ($\sqrt{Q^2+U^2+V^2}/I$) in the \ion{Fe}{1} 6302.5 \AA\ line (right panel).  
The analysis described in the rest of this section was performed on sub-selections of spatial regions based on the continuum intensity and polarization fraction shown in these maps.  
The orange contour in the left panel shows the continuum intensity at a value of $0.454 I_c$ which was used to define the sunspot umbra.  The gray contours in the right panel show the 5\% level of fractional polarization in the \ion{Fe}{1} line.  
An example of the original Stokes spectra after normalization and removal of the residual continuum in Stokes V is shown in the top four panels of Figure \ref{fig:spectra}.  These come from the region indicated by the dashed red line in Figure \ref{fig:maps}.

\subsection{Generation of Polarization Cross-Talk}
To test our ad hoc calibration we took the map of Stokes vectors as measured by SOT/SP as a ground-truth reference and then applied some artificial cross-talk which was then removed by our method.  
To generate the cross-talk we used a realistic \muller matrix based on a preliminary model of the DKIST telescope optics including M1 through M6 \citep[][Harrington et al. 2022, submitted to the DKIST special issue of {\it Solar Physics}]{jatis7}.  
We use this model because we have access to it, have good knowledge of how it was produced, and understand that it represents a realistic optical system.
The diattenuation in the model was artificially enhanced by a constant factor so that it could be easily seen.  
The cross-talk \muller matrix was then applied to the data to generate a corrupted Stokes vector.
For an arbitrary date and time during the local observing day we determined the azimuth and elevation of the Sun in the sky above DKIST based on the solar ephemeris (azimuth: $87.0^\circ$, elevation: $49.5^\circ$).  
The model \muller matrix was then calculated for this telescope azimuth and elevation, with a coud\'e rotator angle of $-30^\circ$, and at a wavelength of 6302 \AA.  This produced a \muller matrix with the following elements:
\begin{equation}\label{eqn:morig}
    \mat{M}_{Orig} = \left[ {\begin{array}{rrrr}
     1 .   &  0.007 &  0.053 & -0.020 \\
    -0.033 &  0.728 & -0.592 &  0.344 \\
     0.047 &  0.674 &  0.704 & -0.220 \\
     0.002 & -0.112 &  0.392 &  0.911 \\
    \end{array} } \right]
\end{equation}
The matrix elements have been rounded to three significant digits, although the calculations were carried out using double precision floating point values.  
Henceforth we will call this the ``original'' system \muller matrix.  
We note the decomposition of this matrix using Equation~\ref{eqn:decomp} implies a diattenuation vector of $[-0.033,0.047,0.002]^{T}$ and Poincare sphere rotation angles of $(\phi,\delta,\theta) = (-16.0^\circ,24.1^\circ,57.4^\circ)$.  
The second row of panels in Figure \ref{fig:spectra} shows the corrupted Stokes Q, U, and V spectra along the red dashed line in Figure \ref{fig:maps}, i.e. the original Stokes vector transformed by $\mat{M}_{Orig}$.  The new Stokes I spectrum appears almost identical to the original, so it is not shown.

\subsection{Model-Based Approach}
The minimization against the diattenuation model was applied first as suggested in Section \ref{sec:implementation} to determine the best fit $D$, $\alpha$, and $\beta$ parameters of the model given in Equation \ref{eqn:diattenuator}.  
The minimization was supplied with a subset of the spatial area of the raster.  Only those regions with low polarization signals, outside of the gray contour in Figure \ref{fig:maps}, were used.  
We did not do any special selection based on wavelength, the full spectrum from these regions was supplied to the minimization.  
After the best fit parameters were determined, the diattenuation \muller matrix was calculated according to Equation \ref{eqn:diattenuator}, then it was inverted and applied to the corrupted Stokes vector for each spectral and spatial pixel in the dataset to correct for the cross-talk between Stokes I and the polarized states.

Next we optimized the retardance model with only the $\mat{M}_{R3}$ and $\mat{M}_{R2}$ matrices from Equation \ref{eqn:retarder}, using the minimization criteria in Equation \ref{eqn:min2} to determine the $\delta$ and $\theta$ parameters.  
In this step we used only the regions with high polarization signal within the gray contour shown in the right panel of Figure \ref{fig:maps}.  We again used all pixels in the spectral dimension, no special spectral selection was done.  
The best fit model \muller matrix was calculated using parameters from the minimization, then it was inverted and applied to every spatial location to correct the cross-talk between Stokes Q and U and Stokes V.

After this step we did a comparison against the original Stokes profiles to manually resolve any global (uniformly applied in $x,\ y, \lambda$) sign differences in Stokes Q, U, and V due to the degeneracies in the technique.  We did not attempt to determine the residual rotation of Q and U using the sunspot geometry.  
The \muller matrices from each of the three steps were multiplied together in order to retrieve the system \muller matrix:
\begin{equation}\label{eqn:tw_result}
    \mat{M}_{This Work} =
    \left[ {\begin{array}{rrrr}    
     1.    & 0.022 &  0.049 & -0.019 \\
    -0.033 & 0.528 & -0.776 &  0.342 \\
     0.047 & 0.848 &  0.485 & -0.213 \\
     0.002 & 0.000 &  0.402 &  0.914 \\
    \end{array} } \right].
\end{equation}
The Stokes Q, U, and V components recovered by our model-based technique are shown in the third row of panels in Figure \ref{fig:spectra}.

\subsection{Alternate Approach}
For the purposes of comparison with our model-based approach, we did a determination of the cross-talk terms from Stokes I to Q, U, and V based on the technique of \citet{sanchez92}, and a determination of the cross-talk between Stokes Q and U and Stokes V based on the technique of \citet{kuhn94}.  We denote this technique SL92/K94.

To get the Stokes I to Stokes Q, U, and V cross-talk terms we selected a section of the continuum from the spectrum and took the mean value of $Q/I$, $U/I$, and $V/I$ over all spatial and spectral pixels within the continuum region.  In the style of \citet{elmore10}, we call these coefficients $e$, $f$, and $g$ respectively.  Based on these parameters we constructed the following inverse matrix:
\begin{equation}
    \mat{M}_{SL92}^{-1} = \left[ {\begin{array}{rrrr}
    1 & 0 & 0 & 0 \\
    -e & 1 & 0 & 0 \\
    -f & 0 & 1 & 0 \\
    -g & 0 & 0 & 1 \\
    \end{array} } \right].
\end{equation}
This matrix was used to produce a corrected Stokes vector for the next step.

\begin{figure*}[t]
    \centering
    \includegraphics[width=6.0in]{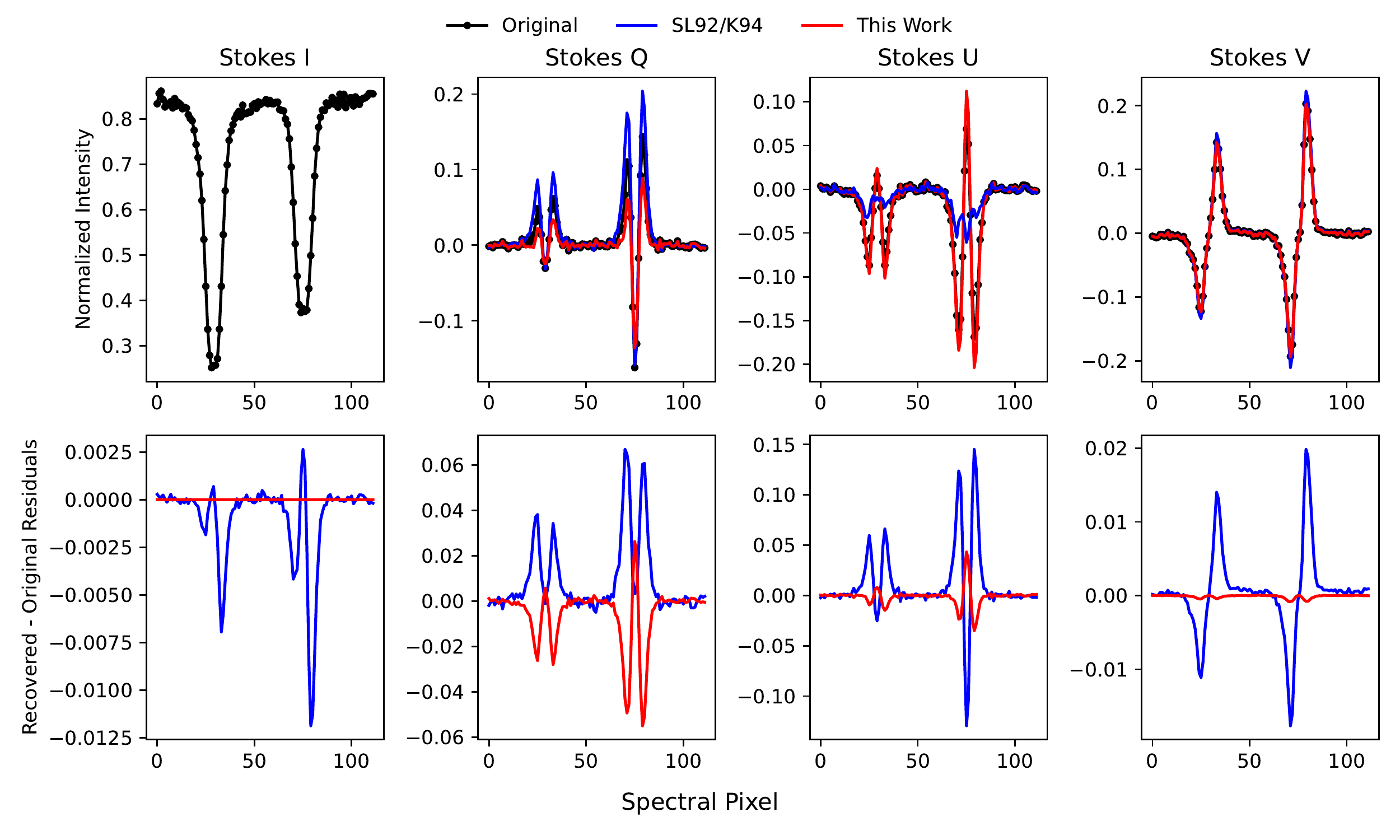}
    \caption{Example Stokes profiles for the position indicated by the circular marker Figure \ref{fig:maps} and the dashed line in Figure \ref{fig:spectra}.  The top row shows the original and retrieved profiles from our method and the Kuhn method.  The lower row of panels shows the difference between the original and recovered profiles.}
    \label{fig:profiles}
\end{figure*}

In the method of \citet{kuhn94}, the cross-talk from Stokes Q and U into Stokes V is determined and removed before the cross-talk from Stokes V into Q and U.  
This technique requires detailed knowledge of the line center and width, so we first determined the Fe I 6302.5 \AA\ line center by taking the center of mass of the polarized line profile, using a spectral selection that excluded the other \ion{Fe}{1} line.  
All profiles were shifted to center them at the same spectral pixel.  The signal in the central Zeeman component was determined for Stokes Q, U, and V by taking the average over the line core, $\pm2$ pixels from the center wavelength.  
A multiple linear regression was used to determine the cross-talk coefficients of Stokes Q and U into Stokes V ($a$ and $b$ in \citet{kuhn94}).  A corrected Stokes V was constructed using the coefficients determined from the linear regression.  
The V to Q and U parameters ($c$ and $d$) were determined by taking the median of $\sum_{xy\lambda} QV / \sum_{xy\lambda} V^2$ and $\sum_{xy\lambda} UV / \sum_{xy\lambda} V^2$ where the sum used the full line profile of the Fe I 6302.5 \AA\ line.  
The coefficients were then used to calculate the inverse matrix according to \citet{kuhn94}:
\begin{equation}
    \mat{M}_{K94}^{-1} =
  \left[ {\begin{array}{cccc}
    1 & 0    & 0    & 0  \\
    0 & 1+ac & cb   & -c \\
    0 & ad   & 1+bd & -d \\
    0 & -a   & -b   & 1  \\
    \end{array} } \right].
\end{equation}
Note that this is actually transposed from the format given in \citet{kuhn94} so that the matrix operates on a column Stokes vector, i.e. $\vec{S}=\mat{M}^{-1} \vec{S}^\prime$.  
This matrix was used to correct the data resulting from the previous step (correction of Stokes I cross-talk).

Similar to the case of our model-based approach, we manually constructed a third matrix to resolve the global sign differences to make the cross-talk corrected spectra match the sign of the original Stokes spectra.
The matrices from these three steps were inverted and multiplied to reconstruct the system matrix:
\begin{equation}\label{eqn:sk_result}
    \mat{M}_{SL92/K94} = 
    \left[ {\begin{array}{rrrr}
     1.    &  0.    &  0.   &  0.    \\
    -0.033 &  1.    &  0.   &  0.314 \\
     0.047 &  0.    &  1.   & -0.192 \\
     0.001 & -0.377 & 0.248 &  0.834 \\
    \end{array} } \right].
\end{equation}
Example Stokes Q, U, and V spectra recovered by the SL92/K94 technique are shown in the bottom row of panels in Figure \ref{fig:spectra}.

\section{Results}\label{sec:results}

\subsection{Recovered Stokes Spectra}
Once again consider Figure \ref{fig:spectra}, which shows example spectra from one step in the SOT/SP raster.  This figure shows the original Stokes spectra (first row), the Stokes spectra after applying the cross-talk \muller matrix (second row), the Stokes spectra recovered by our model-based minimization method (third row), and the Stokes spectra recovered by the SL92/K94 method (fourth row).  There is no apparent difference in Stokes I, so this is only shown once in the top row.

At first glance, the results from our method and the SL92/K94 method look very similar.  Both techniques are able to recover symmetric Stokes Q and U profiles, and anti-symmetric Stokes V profiles. 
There are visible differences in the Stokes Q and U vectors recovered versus the original.  Because we have not applied the final rotation that accounts for the geometric frame of Q and U, each pair of Stokes Q and U components shows a slightly different rotation state of the linear polarization.

There are subtle differences in the recovered Stokes vector that can be seen more easily when inspecting single profiles.
Figure \ref{fig:profiles} shows the Stokes profiles from one spatial position indicated by the red dashed line in Figure \ref{fig:spectra} and the red dot in Figure \ref{fig:maps}.  
The upper set of panels shows the Stokes I, Q, U, and V profiles from the original Stokes vector (black line and data points), the Stokes vector recovered by our method (red line), and the Stokes vector recovered by the SL92/K94 method (blue line).  The lower set of panels shows the Stokes profiles recovered by each method subtracted from the original profiles.  
We can now easily see that the recovered Stokes Q and U profiles are quite different from the originals, but this is expected because there is still a geometric rotation to resolve.  
What is interesting is the Stokes V residual.  At this point, the Stokes V profile should be essentially the same as the input.  Our technique (red line) retrieves the profile of Stokes V at the $10^{-3}$ level with respect to the average quiet-Sun continuum.  The combined SL92/K94 method has a much larger residual at the few percent level (blue line).  The residual in Stokes V from the SL92/K94 method looks like Stokes V.  This method is not failing to remove cross-talk, but it is causing an overall change in the amplitude of the polarized components with respect to the original input.

We can see the problem with the amplitude of the polarized components a little more clearly by comparing the net unsigned polarization for the original Stokes vector and for the Stokes vector recovered by the two techniques over the full dataset.  
We calculate the quantity  $\sum_{xy\lambda}\sqrt{Q^2+U^2+V^2}$ for the original Stokes spectra and for the Stokes spectra recovered by each technique.  For our technique the ratio of total net polarization with respect to the original data is 1.0 to more than five digits of precision.  For the SL92/K94 technique this ratio is 1.0436.  This means there is excess polarization signal with respect to Stokes I, and energy is not being conserved.

\begin{figure*}
    \centering
    \includegraphics[width=7.0in]{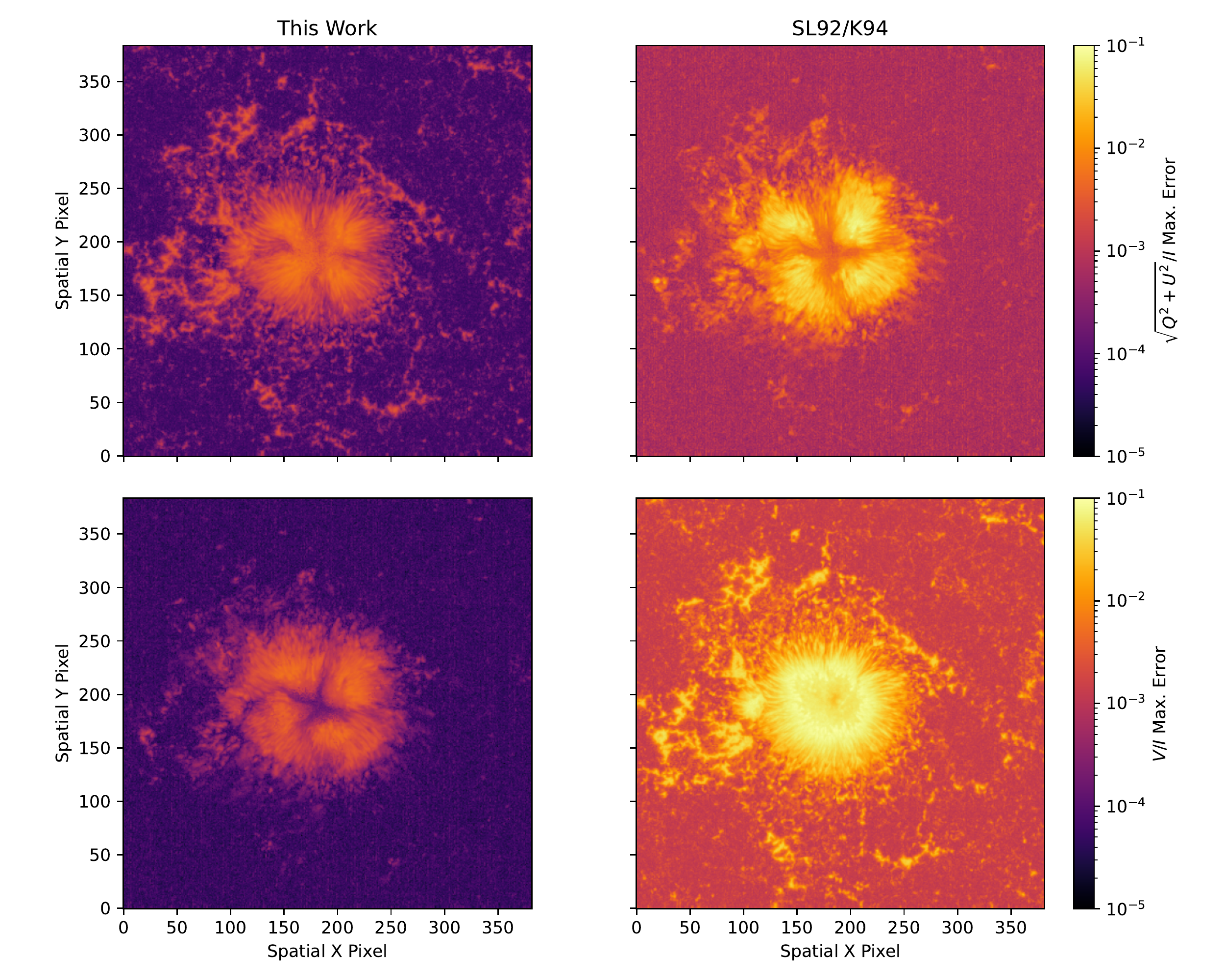}
    \caption{Log scaled maps of the maximum value of the residual signal (recovered - original) in the 6302.5 \AA\ \ion{Fe}{1} line for the combined linear polarization (top panels) and circular polarization (bottom panels) normalized by Stokes I.  The results of the models-based technique from this work (left panels) are compared to the results of the SL92/K94 technique (right panels) for polarization correction.}
    \label{fig:error_maps}
\end{figure*}

The errors in Q and U are due primarily to an unresolved rotation of the linear coordinate system, which has no inherent physical significance.  This is easily seen by looking at the error in the combined linear polarization signal in the recovered data.  
Figure \ref{fig:error_maps} shows the maximum residual of the difference between recovered and original profiles for total fractional linear polarization $\sqrt{Q^2+U^2}/I$ (top row) and circular polarization $\abs{V/I}$ (bottom row); the maximum is determined over wavelength in the 6302.5 \AA\ line \ion{Fe}{1} line and is displayed on a log scale.
For our model-based technique, the maximum error in the combined linear polarization signal is on par with the error in the circular polarization signal at the ~$10^{-3}$ level as a fraction of the Stokes I signal.  
Some level of error in the recovery of the linear and circular polarization states is to be expected due to area asymmetries in the line profiles resulting from physical processes in the line forming region.  There may also be uncorrected polarization in SOT/SP spectra at the $10^{-3}$ level.
In contrast, the SL92/K94 technique (right column) shows much larger residuals in both linear and circular polarization, peaking near $10^{-1}$, and consistently an order of magnitude larger than our method throughout the entire spatial domain.  This difference is due primarily to the amplitude error in their treatment of the total polarization vector.

The spatial distribution of the errors in these maps is worth noting.  For the results of this work, the linear polarization residual shows enhanced signal in the umbra and active network, while the circular polarization residual shows enhanced signal mainly in the sunspot itself.  The residuals in linear polarization look like circular polarization, and the residuals in circular polarization look like linear polarization.  This is the situation we would expect if the minimization was influenced by slight asymmetries in the Stokes profiles from solar sources.  In comparison the SL92/K94 results show the reverse.  The residuals in linear polarization look like linear polarization, and the residuals in circular polarization look like circular polarization.  This is another impact of this technique's failure to recover the amplitude of the components of the Stokes vector properly.

\subsection{Recovered \muller Matrices}
The recovered \muller matrices can also show us how accurate these techniques are.
Consider the resulting \muller matrix from our model-based minimization shown in Equation \ref{eqn:tw_result} and the original system \muller matrix in Equation \ref{eqn:morig}.
This is very similar to $\mat{M}_{Orig}$.  The first column (produced only by the diattenuator model) is highly accurate, as are the terms in the last column (which are related to the cross-talk from Stokes V into other states).  The terms in the middle of the array are somewhat different than in the original \muller matrix.  
If we invert this matrix and apply it to the original we can look at the residual uncorrected terms in the \muller matrix:
\begin{gather}
    \mat{M}_{This Work}^{-1} \mat{M}_{Orig} = \nonumber \\ 
    \left[ {\begin{array}{rrrr}
    1. &  0.    & 0.    &  0.    \\
    0. &  0.959 & 0.285 & -0.005 \\
    0. & -0.285 & 0.959 & -0.006 \\
    0. &  0.003 & 0.007 &  1.    \\
    \end{array} } \right].
\end{gather}
From the symmetry of the center 4 terms of the matrix, we can see that a rotation matrix for an angle of $\cos^{-1}{0.959}= 16.5^\circ$ is necessary for the final correction.  If we apply this final correction to our recovered \muller matrix we end up with a system \muller matrix of:
\begin{gather}
    \mat{M}_{This Work} \mat{M}_{R1}(16.5^\circ) = \nonumber \\ 
    \left[ {\begin{array}{rrrr}
     1.    &  0.007 &  0.053 & -0.019 \\
    -0.033 &  0.727 & -0.594 &  0.342 \\
     0.047 &  0.675 &  0.706 & -0.213 \\
     0.002 & -0.114 &  0.386 &  0.914 \\
    \end{array} } \right]
\end{gather}
which has a residual \muller matrix of:
\begin{gather}
    (\mat{M}_{This Work} \mat{M}_{R1}(16.5^\circ))^{-1} \mat{M}_{Orig} = \nonumber \\
    \left[ {\begin{array}{rrrr}
    1. & 0.    & 0.    &  0.    \\
    0. & 1.    & 0.    & -0.003 \\
    0. & 0.    & 1.    & -0.007 \\
    0. & 0.003 & 0.007 &  1.    \\
    \end{array} } \right]
\end{gather}
when compared to the original.
After the final correction for geometric rotation the only remainder is subtle cross-talk between the linear polarization states and Stokes V at the $10^{-3}$ level.  
As mentioned above, this remainder is likely due to slight area asymmetry in the Stokes profiles due to real physical effects on the Sun.

The matrix produced by the SL92/K94 technique also has a highly accurate first column, but otherwise looks quite different from the original \muller matrix.
We can use the same trick looking at the center 4 terms in the residual to determine final angle correction that is necessary to recover Stokes Q and U.  
\begin{gather}
    \mat{M}_{SL92/K94}^{-1} \mat{M}_{Orig} = \nonumber \\
    \left[ {\begin{array}{rrrr}
    1.    &  0.007 &  0.053 & -0.020 \\
    0.    &  0.729 & -0.588 & -0.001 \\
    0.    &  0.673 &  0.701 & -0.009 \\
    0.001 & -0.004 & -0.004 &  1.095 \\
    \end{array} } \right]
\end{gather}
In this case the necessary angle is $-41.8^\circ$.  This makes the \muller matrix after the rotation:
\begin{gather}
    \mat{M}_{SL92/K94} \mat{M}_{R1}(-41.8^\circ) = \nonumber \\ 
    \left[ \begin{array}{rrrr}
     1.    &  0.    &  0.    &  0.    \\
    -0.033 &  0.746 & -0.666 &  0.314 \\
     0.047 &  0.666 &  0.746 & -0.192 \\
     0.001 & -0.117 &  0.436 &  0.834 \\
     \end{array}  \right]
\end{gather}
And the residual of this from the original is:
\begin{gather}
    (\mat{M}_{SL92/K94} \mat{M}_{R1}(-41.8^\circ))^{-1} \mat{M}_{Orig} = \nonumber \\ 
    \left[ {\begin{array}{rrrr}
    1.    &  0.007 &  0.053 & -0.020 \\
    0.    &  0.992 &  0.028 & -0.007 \\
    0.    &  0.016 &  0.915 & -0.006 \\
    0.001 & -0.004 & -0.004 &  1.095 \\
    \end{array} } \right].
\end{gather}
Ideally this operation should retrieve something that is close to the identity matrix.  While the Stokes Q and U to V cross-talk terms (last row and column) are actually fairly small, the terms along the diagonal are significantly different from 1.
The format of the matrix assumed by the \citet{kuhn94} method is the root of the amplitude error.
It is already apparent that this matrix is not a physically valid \muller matrix because it does not conserve energy, but we further test it realism using the inequalities in \citet{kostinski93} Equations 9 and 11.  Taken together, these inequalities test if a \muller matrix is over-polarizing (i.e. not obeying energy conservation).  The SL92/K94 matrix fails both of these tests, while the original \muller matrix and the one recovered by our technique pass them as expected because these are both non-depolarizing \muller matrices.

\section{Discussion and Conclusions}\label{sec:end}
The primary concept we have put forward is that ad hoc corrections of polarized cross-talk in spectropolarimetric observations can be treated with a physical model composed of a general diattenuator and an elliptical retarder.
This same kind of model is used for modeling telescope optical paths empirically using measurements made with calibration polarizers.
This model is not complicated, and it can be used to accurately correct arbitrary levels of cross-talk on an observation-by-observation basis and gain knowledge about the uncorrected polarization in the telescope and instrument optical system.
Our model is exact for the case of ideal, non-depolarizing optics.
In contrast, previous ad hoc techniques produce a \muller matrix which is an approximation to the non-depolarizing model in the limit of weak polarization.
This approximation does not describe the properties of physical optical elements; i.e., it does not obey Maxwell's equations.  
This is the primary difference between the present work and prior methods and is the reason we are able to obtain higher accuracy.

To find the best fit parameters for the diattenuator and retarder components of the model, we have implemented separate minimization criteria that are appropriate for spectra with an unpolarized continuum containing polarized signals produced by photospheric lines sensitive to the Zeeman effect.  Such lines are commonly observed as a diagnostic of solar magnetic fields.
We applied this technique to a real observation from \textit{Hinode} SOT/SP with known added polarization cross-talk and were able to retrieve the linear and circular polarization intensities with accuracy at the $10^{-3}$ level or lower relative to the quiet-Sun continuum intensity.

Although there may be residual polarization in the SOT/SP spectropolarimetry at the $10^{-3}$ level, asymmetries in the line shapes are likely the limiting factor in accurate recovery of the Stokes vector.
Asymmetries in the line profiles may occur due to line of sight velocity and/or magnetic field gradients across the line formation region or departures from LTE formation \citep{lopez02}.  These are often greater in the sunspot penumbra or along polarity inversion lines \citep[some examples include  ][]{sanchez92,deng10,kaithakkal20}.  Improvements in the accuracy of our method may be possible by avoiding regions known to show asymmetries.

Because of the symmetry and anti-symmetry properties of the polarization in the Zeeman-split line profiles, there is a sign ambiguity in Stokes V, and an ambiguity in the geometric frame of Stokes Q and U, that remain.  These need to be resolved after the minimization technique is applied in order to get the proper system \muller matrix and recover the Stokes vector produced by the Sun.
Any ambiguities of this kind could also be resolved in the magnetic field geometry after a spectropolarimetric inversion for the magnetic field has been done.

For the problem of cross-talk in spectropolatrimetry of the solar photosphere, this technique is very simple to apply and does not require any detailed selection based on wavelength.
The minimization criteria are sufficiently general that they should  apply to any polarized line signature with zero net circular polarization when summed over wavelength, and Stokes Q and U profiles that have an appearance different than Stokes V.
This would include non-triplet Zeeman lines, lines produced through the Paschen-Bach effect, and complex blends of many lines.  
The minimization does require strong polarized signals in Stokes Q, U, and V states.
A large number of profiles with diverse signals might improve the accuracy of this technique, but it is not actually required for the minimization to work.  Even one set of profiles with sufficient signal should be enough.

We have suggested one minimization scheme that works for strong signals, such as those produced by active regions, but alternative minimization schemes would be necessary to apply this technique to other situations.
Quiet-Sun magnetic fields have very weak Q and U signals, and in this case the $QV$ and $UV$ terms in our minimization Equation \ref{eqn:min2} would fail to provide a strong constraint on the model.
It might be possible to pair the cross-talk modeling with a simple spectropolarmetric inversion, using the weak-field approximation or Milne-Eddington fit, to minimize the observed profiles against model profiles.

A minimization is not actually necessary to determine the diattenuation terms for this application.  Taking a mean or median of the quantities $Q/I$, $U/I$, and $V/I$ over regions of quiet-Sun continuum works quite well to retrieve the correct $d_H$, $d_{45}$, and $d_R$ terms respectively.  The significant advantage of our technique over \citet{sanchez92} is in using these terms to construct the proper model for Stokes I cross-talk given by the diattenuation matrix in Equation \ref{eqn:diattenuator}.

For our cross-talk removal technique, and those of previous authors, any real continuum polarization would be removed along with cross-talk from Stokes I due to diattenuation.  While the spectral profiles might be adequately corrected, the resulting \muller matrix will not be correct.  Because the continuum polarization from scattering is fairly well understood from a theoretical standpoint, it might be possible to disentangle the diattenuation by modeling the linear polarization magnitude and direction based on the location of the source region on the Sun and adding that to the minimizing functional.  In that case the linear polarization direction is known, so modeling it along with the polarization cross-talk could also solve the unknown geometric rotation needed to retrieve Stokes Q and U.

The assumptions of uniform polarization response over wavelength and field of view made in Section \ref{sec:implementation} need to be evaluated on a case by case basis for the telescope and instrument system where this technique is applied.
Non-uniformities, particularly from incidence angle variation with field angle, can introduce significant spatial and wavelength dependent effects \citep[e.g.][Section 6.1, Figure 19]{jatis1}. 
Some non-uniformity of the polarization response might be mitigated with an appropriate calibration strategy for the instrument, using a spatially or wavelength dependent demodulation.
If non-uniformities still persist, our technique could be applied to diagnose and correct them by using spatial or spectral sub-regions of the data assuming sufficient polarized signal is present in these smaller regions.

Our model-based approach has a significant advantage over previous ad hoc techniques of instrumental polarization correction in that it can be applied to any general optical system with arbitrary amounts of cross-talk, provided that it is non-depolarizing.  
Previously published techniques were only intended to be applied in situations where optics are weakly polarizing, or the majority of instrumental cross-talk has been removed by other means.  
These methods fail to recognize that cross-talk to and from various polarization states is physically linked, and they apply the cross-talk correction using an approximation to the \muller matrix that does not describe a physical optical system.  
In regimes where cross-talk is large, these techniques produce increasingly unphysical Stokes vectors and transformation matrices that do not conserve energy or the polarized signal.  
Using incorrect Stokes vector amplitudes in spectropolarimetric inversions could have far ranging effects on interpretation of the magnetic field strength and direction, and the magnetic field filling factor, especially where the polarized signals are large.

\begin{acknowledgments}
The National Solar Observatory (NSO) is managed by the Association of Universities for Research in Astronomy, Inc., and is funded by the National Science Foundation.  Any opinions, findings and conclusions or recommendations expressed in this publication are those of the authors and do not necessarily reﬂect the views of the National Science Foundation or the Association of Universities for Research in Astronomy, Inc.

Hinode is a Japanese mission developed and launched by ISAS/JAXA, collaborating with NAOJ as a domestic partner, NASA and STFC (UK) as international partners. Scientific operation of the Hinode mission is conducted by the Hinode science team organized at ISAS/JAXA. This team mainly consists of scientists from institutes in the partner countries. Support for the post-launch operation is provided by JAXA and NAOJ(Japan), STFC (U.K.), NASA, ESA, and NSC (Norway).

This research has made use of NASA’s Astrophysics Data System.
\end{acknowledgments}

\vspace{5mm}
\facilities{Hinode(SOT/SP)}


\bibliography{bibliography}{}
\bibliographystyle{aasjournal}

\end{document}